%% file: main.tex
\documentclass[fleqn,usenatbib]{mnras}

\usepackage{graphicx}
\graphicspath{{figures}}

\usepackage{float}
\usepackage{booktabs}

\usepackage{amsmath}
\usepackage{amssymb}

\usepackage{xcolor}
\usepackage{orcidlink}

\usepackage{newtxtext,newtxmath}

\usepackage[normalem]{ulem}

\newcommand{\raisetarget}[2]
{\bgroup
  \sbox0{#2}%
  \raisebox{\ht0}{\hypertarget{#1}{}}\usebox0%
\egroup}

\title[TESS Giants Transiting Giants V]{TESS Giants Transiting Giants V - Two hot Jupiters orbiting red-giant hosts}

\author[F. Pereira et al.]{
	Filipe Pereira \orcidlink{0000-0002-2157-7146} 
		\hyperlink{Affilliation_1}{$^{1}$}$^{,}$\hyperlink{Affilliation_2}{$^{2}$}$^{,}$$^{\thanks{\href{mailto:filipe.pereira@astro.up.pt}{filipe.pereira@astro.up.pt}}}$,
	Samuel K. Grunblatt \orcidlink{0000-0003-4976-9980}
		\hyperlink{Affilliation_3}{$^{3}$}$^{,}$\hyperlink{Affilliation_4}{$^{4}$}$^{,}$\hyperlink{Affilliation_6}{$^{6}$}, 
	Angelica Psaridi \orcidlink{0000-0002-4797-2419} 
		\hyperlink{Affilliation_5}{$^{5}$},
	Tiago L. Campante \orcidlink{0000-0002-4588-5389}
		\hyperlink{Affilliation_1}{$^{1}$}$^{,}$\hyperlink{Affilliation_2}{$^{2}$},
	\newauthor 
	Margarida S. Cunha \orcidlink{0000-0001-8237-7343}
		\hyperlink{Affilliation_1}{$^{1}$}$^{,}$\hyperlink{Affilliation_2}{$^{2}$},
	Nuno C. Santos \orcidlink{0000-0003-4422-2919}
		\hyperlink{Affilliation_1}{$^{1}$}$^{,}$\hyperlink{Affilliation_2}{$^{2}$},
	Diego Bossini \orcidlink{0000-0002-9480-8400}
		\hyperlink{Affilliation_1}{$^{1}$},
	Daniel Thorngren \orcidlink{0000-0002-5113-8558}
		\hyperlink{Affilliation_6}{$^{6}$},
	\newauthor 
	Coel Hellier \orcidlink{0000-0002-3439-1439}
		\hyperlink{Affilliation_7}{$^{7}$},
	François Bouchy \orcidlink{0000-0002-7613-393X}
		\hyperlink{Affilliation_5}{$^{5}$},
	Monika Lendl \orcidlink{0000-0001-9699-1459}
		\hyperlink{Affilliation_5}{$^{5}$},
	Dany Mounzer \orcidlink{0000-0002-8070-2058}
		\hyperlink{Affilliation_5}{$^{5}$},
	Stéphane Udry \orcidlink{0000-0001-7576-6236}
		\hyperlink{Affilliation_5}{$^{5}$},
	\newauthor
	Corey Beard \orcidlink{0000-0001-7708-2364}
		\hyperlink{Affilliation_8}{$^{8}$}$^{,}$$^{\thanks{NASA FINESST Fellow}}$,
	Casey L. Brinkman \orcidlink{0000-0002-4480-310X}
		\hyperlink{Affilliation_9}{$^{9}$},
	Howard Isaacson \orcidlink{0000-0002-0531-1073}
		\hyperlink{Affilliation_10}{$^{10}$}$^{,}$\hyperlink{Affilliation_11}{$^{11}$},
	Samuel N. Quinn \orcidlink{0000-0002-8964-8377}
		\hyperlink{Affilliation_12}{$^{12}$},
	\newauthor
	Dakotah Tyler \orcidlink{0000-0003-0298-4667}
		\hyperlink{Affilliation_13}{$^{13}$},
	George Zhou \orcidlink{0000-0002-4891-3517}
		\hyperlink{Affilliation_11}{$^{11}$},
	Steve B. Howell \orcidlink{0000-0002-2532-2853}
		\hyperlink{Affilliation_14}{$^{14}$},
	Andrew W. Howard \orcidlink{0000-0001-8638-0320}
		\hyperlink{Affilliation_15}{$^{15}$},
	\newauthor
	Jon M. Jenkins \orcidlink{0000-0002-4715-9460}
		\hyperlink{Affilliation_14}{$^{14}$},
	Sara Seager \orcidlink{0000-0002-6892-6948}
		\hyperlink{Affilliation_16}{$^{16}$}$^{,}$\hyperlink{Affilliation_17}{$^{17}$}$^{,}$\hyperlink{Affilliation_18}{$^{18}$},
	Roland K. Vanderspek \orcidlink{0000-0001-6763-6562}
		\hyperlink{Affilliation_16}{$^{16}$},
	Joshua N. Winn \orcidlink{0000-0002-4265-047X}
		\hyperlink{Affilliation_19}{$^{19}$},
	\newauthor
	Nicholas Saunders \orcidlink{0000-0003-2657-3889}
		\hyperlink{Affilliation_9}{$^{9}$},
	Daniel Huber \orcidlink{0000-0001-8832-4488}
		\hyperlink{Affilliation_9}{$^{9}$}$^{,}$\hyperlink{Affilliation_11}{$^{20}$}
	\\
	\raisetarget{Affilliation_1}{$^{1}$}Instituto de Astrofísica e Ciências do Espaço, Universidade do Porto, CAUP, Rua das Estrelas, P-4150-762 Porto, Portugal \\
	\raisetarget{Affilliation_2}{$^{2}$}Departamento de Física e Astronomia, Faculdade de Ciências da Universidade do Porto, Rua do Campo Alegre, P-4169-007 Porto, Portugal \\
	\raisetarget{Affilliation_3}{$^{3}$}American Museum of Natural History, 200 Central Park West, Manhattan, NY 10024, USA \\
	\raisetarget{Affilliation_4}{$^{4}$}Center for Computational Astrophysics, Flatiron Institute, 162 5th Avenue, Manhattan, NY 10010, USA \\
	\raisetarget{Affilliation_5}{$^{5}$}Observatoire Astronomique de l'Université de Genève, Chemin Pegasi 51, 1290 Versoix, Switzerland \\
	\raisetarget{Affilliation_6}{$^{6}$}Department of Physics \& Astronomy, Johns Hopkins University, Baltimore, MD, USA \\
	\raisetarget{Affilliation_7}{$^{7}$}Astrophysics Group, Keele University, Staffordshire ST5 5BG, UK \\
	\raisetarget{Affilliation_8}{$^{8}$}Department of Physics \& Astronomy, University of California, Irvine, Irvine, CA 92697, USA \\
	\raisetarget{Affilliation_9}{$^{9}$}Institute for Astronomy, University of Hawai'i, 2680 Woodlawn Drive, Honolulu, HI 96822 USA \\
	\raisetarget{Affilliation_10}{$^{10}$}Institute for Astronomy, University of California Berkeley, Berkeley, CA 94720, USA \\
	\raisetarget{Affilliation_11}{$^{11}$}University of Southern Queensland, Centre for Astrophysics, West Street, Toowoomba, QLD 4350, Australia \\
	\raisetarget{Affilliation_12}{$^{12}$}Center for Astrophysics | Harvard \& Smithsonian, 60 Garden Street, Cambridge, MA 02138, USA \\
	\raisetarget{Affilliation_13}{$^{13}$}Department of Physics \& Astronomy, University of California Los Angeles, Los Angeles, CA 90095, USA \\
	\raisetarget{Affilliation_14}{$^{14}$}NASA Ames Research Center, Moffett Field, CA 94035, USA \\
	\raisetarget{Affilliation_15}{$^{15}$}Department of Astronomy, California Institute of Technology, Pasadena, CA 91125, USA \\
	\raisetarget{Affilliation_16}{$^{16}$}Department of Physics and Kavli Institute for Astrophysics and Space Research, Massachusetts Institute of Technology, Cambridge, MA 02139, USA \\
	\raisetarget{Affilliation_17}{$^{17}$}Department of Earth, Atmospheric and Planetary Sciences, Massachusetts Institute of Technology, Cambridge, MA 02139, USA \\
	\raisetarget{Affilliation_18}{$^{18}$}Department of Aeronautics and Astronautics, MIT, 77 Massachusetts Avenue, Cambridge, MA 02139, USA \\
	\raisetarget{Affilliation_19}{$^{19}$}Department of Astrophysical Sciences, Princeton University, Princeton, NJ 08544, USA \\
	\raisetarget{Affilliation_19}{$^{20}$}Sydney Institute for Astronomy (SIfA), School of Physics, University of Sydney, NSW 2006, Australia
}

\date{Accepted XXX. Received YYY; in original form ZZZ}
\pubyear{2023}

\begin{document}
\label{firstpage}
\pagerange{\pageref{firstpage}--\pageref{lastpage}}
\maketitle


\begin{abstract}
	In this work we present the discovery and confirmation of two hot Jupiters orbiting red-giant stars, TOI-4377 b and TOI-4551 b, observed by TESS in the southern ecliptic hemisphere and later followed-up with radial-velocity (RV) observations.
	For TOI-4377 b we report a mass of $0.957^{+0.089}_{-0.087} \ M_\mathrm{J}$ and a inflated radius of $1.348 \pm 0.081 \ R_\mathrm{J}$ orbiting an evolved intermediate-mass star ($1.36 \ \mathrm{M}_\odot$, $3.52 \ \mathrm{R}_\odot$; TIC 394918211) on a period of of $4.378$ days.
	For TOI-4551 b we report a mass of $1.49 \pm 0.13 \ M_\mathrm{J}$ and a radius that is not obviously inflated of $1.058^{+0.110}_{-0.062} \ R_\mathrm{J}$, also orbiting an evolved intermediate-mass star ($1.31 \ \mathrm{M}_\odot$, $3.55 \ \mathrm{R}_\odot$; TIC 204650483) on a period of $9.956$ days.
	We place both planets in context of known systems with hot Jupiters orbiting evolved hosts, and note that both planets follow the observed trend of the known stellar incident flux-planetary radius relation observed for these short-period giants.
	Additionally, we produce planetary interior models to estimate the heating efficiency with which stellar incident flux is deposited in the planet's interior, estimating values of $1.91 \pm 0.48\%$ and $2.19 \pm 0.45\%$ for TOI-4377 b and TOI-4551 b respectively.
	These values are in line with the known population of hot Jupiters, including hot Jupiters orbiting main sequence hosts, which suggests that the radii of our planets have reinflated in step with their parent star's brightening as they evolved into the post-main-sequence.
	Finally, we evaluate the potential to observe orbital decay in both systems.
\end{abstract}



\section{Introduction}
\label{sec:introduction}

The occurrence rate of giant planets orbiting main sequence and evolved stars has been estimated to be the same, $\sim 10\%$ \citep{Johnson_2007a,Cumming_2008,Mayor_2011,Wittenmyer_2020}, with these estimates obtained from RV surveys, given the long periods of the planets.
Looking only at hot Jupiters, radial-velocity (RV) surveys determine values of their occurrence rate around main sequence stars of $\sim 1-1.5\%$ \citep{Cumming_2008,Mayor_2011}, whilst photometric observations estimate the values to be $\sim 0.5\%$ \citep{Howard_2012,Fressin_2013,Petigura_2018a}.
Some possible explanations put forward for these differences between observational methods include different metallicities between sample \citep{Howard_2012,Wright_2012} and stellar multiplicity \citep{Wang_2015a}.

Looking at hot Jupiters orbiting evolved stars, the situation is less clear.
Through RV surveys, with which the majority of known planets orbiting evolved stars have been identified \citep{Johnson_2010b,Reffert_2015}, there was a dearth of short-period planets found, especially hot Jupiters \citep{Bowler_2010,Johnson_2010a,Reffert_2015,Ottoni_2022}.
These results hinted at possible differences between the populations of hot Jupiters around main sequence and evolved hosts, with planetary spiralling and eventual engulfment being suggested as a mechanism for the depletion of these planets around evolved hosts \citep{Villaver_2009,Schlaufman_2013}.

The discovery of giant exoplanets transiting such evolved hosts is more recent, driven by photometric space missions such as \textit{Kepler} \citep{Huber_2013a,Lillo-Box_2014,Barclay_2015,Quinn_2015,Chontos_2019} and \textit{K2} \citep{VanEylen_2016a,Grunblatt_2016,Grunblatt_2017,Grunblatt_2019}.
These transiting systems were found on shorter periods than their RV counterparts, with its population mostly consisting of hot Jupiters. 
Subsequent studies of these close-in giant planets orbiting \textit{K2} stars revealed that their occurrence rate is $0.37^{+0.29}_{-0.09}\%$ \citep{Grunblatt_2019,Temmink_2023}, mostly in line with the results from main sequence stars.
More precise estimates of the planet occurrence rate of evolved stars will reveal whether late-stage stellar evolution has a significant impact on planet demographics, as has been predicted by evolutionary theory \citep{Schlaufman_2013,Villaver_2014}.

Following up on \textit{Kepler} and \textit{K2}, the Transiting Exoplanet Survey Satellite \citep[TESS;][]{Ricker_2015} traded length of observations for sky coverage, with the 2-year primary mission having observed close to the entire sky for at least one month.
\cite{Campante_2016a} predicted that up to 200 low-luminosity red-giant host stars with asteroseismic oscillations could be detected with TESS.
Currently, TESS has already led to the discovery of a handful of planets orbiting evolved stars.
TOI-197b was the first planetary detection orbiting an evolved star, using TESS data, and also included an asteroseismic detection for the host \citep{Huber_2019b}.
In the meantime, multiple additional transiting giant planets have been found orbiting evolved stars, both subgiants and red giants \citep{Nielsen_2019,Rodriguez_2021,Khandelwal_2022,Saunders_2022,Grunblatt_2022}.
Planets in evolved systems detected by TESS have begun to reveal new features of planetary demographics as a function of evolutionary state \citep{Grunblatt_2023}, and additional transiting planet discoveries will help to determine the significance of these features.

In this article we report the discovery and characterization of two newly discovered short-orbit giant planets, TOI-4377 b and TOI-4551 b, both of which orbit red-giant stars observed by \textit{TESS}.
In Section~\ref{sec:observational_data} we outline all data acquired and utilized in the discovery and characterization of these two planetary systems.
Then, Section~\ref{sec:host_characterization} describes the characterization of the host stars and Section~\ref{sec:planet_characterization} the characterization of the newly discovered planets.
In Section~\ref{sec:discussion} we explore some open questions in the literature related to hot Jupiters and discuss how these two planets fit in with the existing population of close-in giant planets orbiting evolved stars.
Finally, a summary of this work's results and some conclusions are presented in Section~\ref{sec:conclusions}.

\section{Observational Data}
\label{sec:observational_data}

\subsection{\textit{TESS} Photometry}
\label{sec:tess_data}

TOI-4377 b (TIC 394918211) and TOI-4551 b (TIC 204650483), the two discovered planets introduced in this article, were both identified as planet candidates through a survey searching for giant planets orbiting bright, low-luminosity red-giant branch stars (LLRGBs).
The reason for selecting this population of stars was twofold.
Firstly, to try and constrain our stellar sample so that the frequency of maximum oscillation power, $\nu_\mathrm{max}$, was lower than the Nyquist frequency of 30-min cadence TESS data (making stellar oscillations potentially detectable; see Section~\ref{sec:host_astero}).
And secondly, so that Jupiter-sized planets transiting their hosts could be visible in TESS photometric data.

Using available information from the TESS Input Catalog \citep[TIC;][]{Stassun_2019}, we defined empirical cuts on color, magnitude and Gaia parallax to select only the giant stars from the catalog.
To further distinguish LLRGB stars from the remaining giant sample, we initially aimed at restricting the radii of stars from 3 R$_\odot$ to 8 R$_\odot$.
These radii were determined using the Stefan-Boltzmann relation, with Gaia parallaxes used to estimate distances, and the stellar effective temperatures obtained from spectroscopy (when available) or derredened colors (see section 2.3.5 of \citealt{Stassun_2019} for more details).
However, to ensure no true giants were lost due to uncertainties in these TIC radii, we assumed a conservative estimate of their systematic uncertainties of about 15\% \citep{Tayar_2022} and expanded our criterion to stellar radii between 2.5 R$_\odot$ and 10 R$_\odot$.
Finally, we considered only brighter stars in the southern ecliptic hemisphere, restricting our targets to TESS magnitudes below 10.

All targets from the sample observed by TESS were put through a custom-made data processing pipeline comprised mostly of open-source software.
This pipeline extracts and corrects light curves from TESS full-frame images (FFIs)\footnote{TESS FFIs were produced by the TESS Science Processing Operations Center at NASA Ames Research Center \citep{Jenkins_2016}.}, searches for transits and performs statistical and astrophysical false-positive validation of transits, identifying and ranking promising planet candidates.
A full description of the pipeline and open-source software used in it is described in detail in \cite{Pereira_2022}.
Additionally, all promising targets were also analysed with the \textsc{giants} pipeline described in \cite{Saunders_2022}, which has been used to confirm the discovery of giant planets orbiting giant stars with TESS, with promising results \citep{Saunders_2022,Grunblatt_2022}.
This cross evaluation of candidates improved the chances that follow-up time was dedicated to the targets more likely to host giant planets.

In addition to the target search described above, additional targets were independently identified around fainter evolved stars using the \textsc{giants} pipeline and then verified using our independent pipeline.
TOI-4377 (TIC 394918211), the host star of TOI-4377 b, is one such case, as the star has a TESS magnitude of 10.79.

\begin{figure*}
	\centering
	\includegraphics[width=\textwidth]{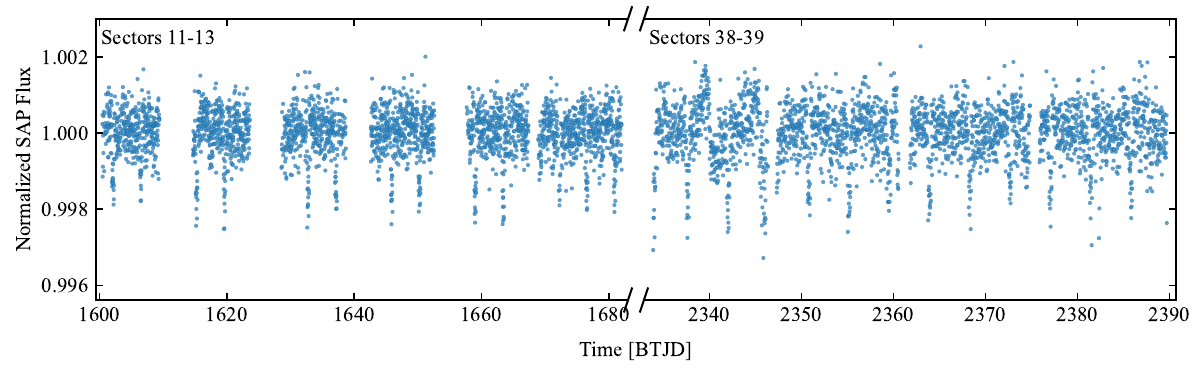}
	\includegraphics[width=\textwidth]{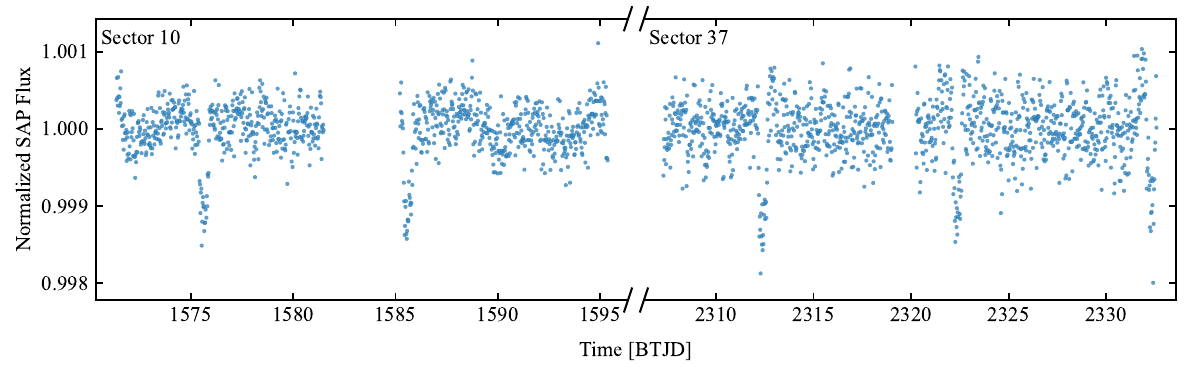}
	\caption{\textit{Top}: Light curve produced for TOI-4377. 
	The left section corresponds to Sectors 11 to 13, observed in 30-min cadence during the first year of TESS.
	The one on the right corresponds to Sectors 38 and 39 from TESS's extended mission, with the measurements, observed on a cadence of 10 minutes, binned to 30 minutes to match the first year sectors.
	\textit{Bottom}: Light curve produced for TOI-4551. 
	The left section corresponds to Sector 10, observed in 30-min cadence during the first year of TESS.
	The one on the right corresponds to Sector 37 from TESS's extended mission, again with the measurements, observed on a cadence of 10 minutes, binned to 30 minutes to match the first year sectors.
	BTJD is the Barycentric TESS Julian Date, defined as the Barycentric Julian Date (BJD) - 2457000.}
	\label{fig:lc_data}
\end{figure*}

TOI-4377 was observed for three sectors in the initial year of TESS (Sectors 11 to 13), with two additional sectors of observations during TESS's extended mission (Sectors 38 and 39), which were observed with a higher cadence (10-min) and binned to 30-min observations in this work, to match the nominal mission's observation cadence.
TOI-4551 was observed for two sectors, Sector 10 in the first year of TESS and Sector 37 in the extended mission, with the observations from the extended mission's sector having also been binned to 30-min cadence in this work.
The light curves produced by our pipeline for TOI-4377 and TOI-4551 are shown on the top and bottom panels of Figure~\ref{fig:lc_data}, respectively.

The transit signatures of both targets also passed all the tests in Data Validation, as presented in the TESS Science Processing Operations Center (SPOC) DV reports \citep{Jenkins_2016,Twicken_2018,Li_2019} on MAST (Mikulski Archive for Space Telescopes).
Both targets were made TOIs by the TESS team on Dec 14, 2022.


\subsection{Spectroscopic Follow-up}
\label{sec:rv_data}

RV follow-up observations were carried out for both planet candidates.
For TOI-4377, 20 spectra were obtained between April 2021 and June 2021 using the high-resolution CORALIE spectrograph, located at ESO's La Silla Observatory on the Swiss 1.2-m Leonhard Euler Telescope \citep{Queloz_2001}.
CORALIE has a resolution of $R \approx 60 \, 000$ and is fed by a $2 \arcsec$ fiber \citep{Segrasan_2010}. The exposure times were set at 1800 s, resulting in spectra with a signal-to-noise ratio per resolution element (S/N) ranging from 8 to 19. We derived the RV of each epoch by cross-correlating the spectrum with a K5 binary mask \citep{Baranne_1996}.
An additional 6 spectra were obtained between May 2021 and June 2021  using the CHIRON spectrograph ($R \approx 80 \, 000$) at the 1.5-meter SMARTS telescope of the Cerro Tololo Inter-American Observatory (CTIO) in Chile \citep{Tokovinin_2013}.
CHIRON uses a ThAr lamp as a wavelength reference and the RVs were derived from each spectrum following \cite{Zhou_2020}.

For TOI-4551, 15 spectra were obtained between May 2021 and August 2021 using the CHIRON spectrograph using the same data processing treatment as previously mentioned. 
An additional eight spectra were also obtained between June 2021 and June 2022, with the HIRES spectrograph ($R \approx 55 \, 000$) at the Keck-I telescope of the W. M. Keck Observatory in Maunakea, Hawaii \citep{Vogt_1994}.
The HIRES spectra were obtained and processed using standard procedures of the California Planet Search \citep{Howard_2010}, including the use of an iodine cell as a wavelength reference and RV determination using the method of \citep{Butler_1996b}.

Tables~\ref{tab:tic394_rv_obs} and \ref{tab:tic204_rv_obs} show all RV measurements, along with their observation times and corresponding uncertainties for TOI-4377 and TOI-4551, respectively.

\subsection{Ground-based Imaging}
\label{sec:ground_imaging}

\begin{figure}
	\centering
	\includegraphics[width=\columnwidth]{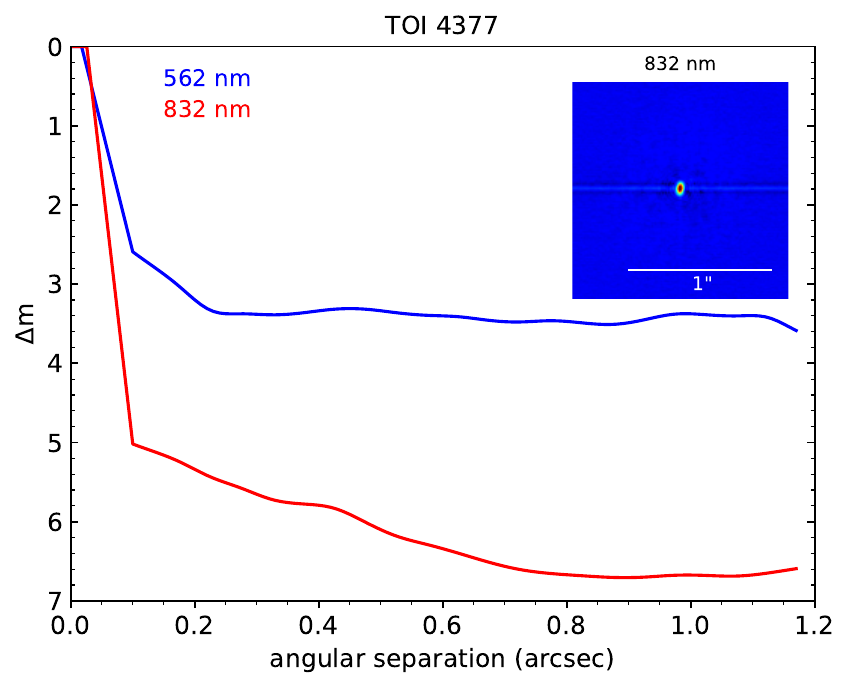}
	\includegraphics[width=\columnwidth]{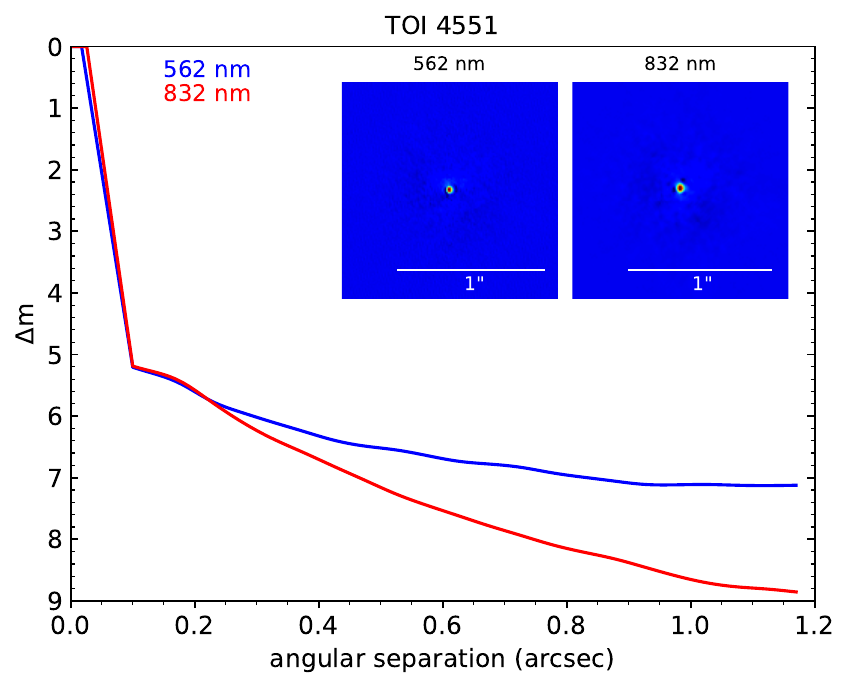}
	\caption{5-$\sigma$ contrast curves of TOI-4377 (top) and TOI-4551 (bottom) for the 562 nm and 832 nm bands, based on speckle imaging obtained with the Zorro speckle instrument.
	The reconstructed speckle images are shown in the insets.
	Both stars show no close companions to within the angular and magnitude contrast limits achieved.}
	\label{fig:speckle_imaging}
\end{figure}

Close stellar companions (bound or line of sight) can confound exoplanet discoveries in a number of ways. 
The detected transit signal might be a false positive due to a background eclipsing binary and even real planet discoveries will yield incorrect stellar and exoplanet parameters if a close companion exists and is unaccounted for \citep{Ciardi_2015}.
Given that nearly one-half of solar-like stars are in binary or multiple star systems \citep[e.g. ][]{Matson_2018}, high-resolution imaging provides validation and characterization information for an exoplanet system and crucial information toward our understanding of exoplanetary formation, dynamics and evolution \citep{Howell_2021}.

TOI-4377 and TOI-4551 were observed on 2022 May 17 UT and 2022 May 18 UT, respectively, using the Zorro speckle instrument on the Gemini South 8-m telescope \citep{Scott_2021,Howell_2022}.
Zorro provides simultaneous speckle imaging in two bands (562nm and 832 nm) with output data products including a reconstructed image with robust contrast limits. 
We obtained eight sets of $1000 \times 0.06$ second images for TOI-4377 and four sets for TOI-4551.
Both data sets were processed in our standard reduction pipeline \citep[see ][]{Howell_2011}.
Figure~\ref{fig:speckle_imaging} shows our 5-$\sigma$ contrast curves for each filter observation and the 832 nm reconstructed speckle image.
The bar seen in the image of TOI-4377 is an artifact of the data reduction processing, likely due to a cosmic ray which was not completely removed.
We find that both stars show no close companions to within the angular (8-m telescope diffraction limit (20 mas) out to 1.2”) and magnitude contrast limits achieved.
At the distance of TOI-4377 ($\mathrm{d} = 455.8$ pc) and TOI-4551 ($\mathrm{d} = 216$ pc) these angular limits correspond to spatial limits of 9 to 547 AU and 4 to 259 AU, respectively.

\subsection{Stellar Rotation}
\label{sec:stellar_rotation}

Based on WASP \citep[Wide Angle Search for Planets;][]{Butters_2010} data (48 000 data points from 2007 to 2014), we find that, for TOI-4551, there is an upper limit of only about 1 mmag on any rotational modulation (between $\sim$3 days and $\sim$80 days).
This points to stellar rotation not being responsible for the periodic variation which we attribute to the discovered planet orbiting this star.
No WASP data are available for TOI-4377.

\section{Host Star Characterization}
\label{sec:host_characterization}

\subsection{Spectroscopy}
\label{sec:host_spec}

From the spectra obtained for each of the targets, we derived stellar atmospheric parameters.

For TOI-4377, available CORALIE spectra were co-added and the atmospheric parameters were derived following the procedure of \cite{Santos_2013,Sousa_2021}.
The values adopted for the effective temperature $T_\mathrm{eff}$, surface gravity $\log g$ and metallicity $\mathrm{[Fe/H]}$ are shown in the middle column of Table~\ref{tab:stellar_properties}.
Atmospheric parameters were also determined using the CHIRON spectra available for this target, as per \cite{Zhou_2020}, and were found to be in agreement with the adopted ones, with $T_\textrm{eff}$ and $\mathrm{[Fe/H]}$ found within 1-$\sigma$ and $\log g$ within 2-$\sigma$.

For TOI-4551, the values adopted were determined using the HIRES spectra and SpecMatch \citep{Petigura_2015}, and are shown in the rightmost column of Table~\ref{tab:stellar_properties}.
Similarly to the previous target, atmospheric parameters were also determined using CHIRON spectra and following \cite{Zhou_2020}, and were found to be in agreement within 1-$\sigma$ in $T_\textrm{eff}$ and $\mathrm{[Fe/H]}$ and within 2-$\sigma$ in $\log g$.

\input{tables/stellar_properties.tex}

Using the atmospheric parameters determined for each star, we then computed stellar masses, radii, densities and ages for both stars, using the Bayesian tool \textsc{PARAM} \citep{daSilva_2006,Rodrigues_2014,Rodrigues_2017}.
\textsc{PARAM} is a grid-based method that matches a set of stellar observables to models in a grid of stellar evolutionary tracks.
In this work, this set of observables consisted of the previously mentioned atmospheric parameters, the Gaia DR3 parallax \citep{GaiaCollaboration_2016,GaiaCollaboration_2022} as well as the $B$, $V$ and 2MASS $JHK_s$ apparent magnitudes provided by the TESS Input Catalog.
We report PARAM constraints on stellar masses, radii, densities, and ages in Table~\ref{tab:stellar_properties}.
Additionally, \textsc{PARAM} can also make use of available asteroseismic data, though in this case, no oscillations were detected in the available TESS photometric data of each target (see Section \ref{sec:host_astero}).

\begin{figure}
	\centering
	\includegraphics[width=\columnwidth]{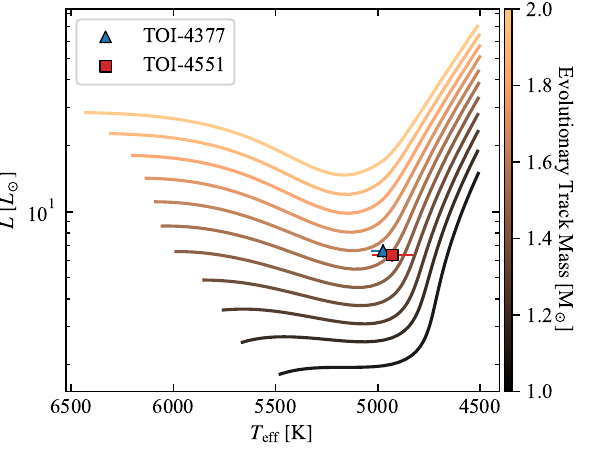}
	\caption{HR diagram with the position of the two hosts, TOI-4377 and TOI-4551, highlighted.
	Also shown are MIST evolutionary tracks depicting the subgiant and red-giant branch evolutionary phases.
	The tracks are color-coded according to their stellar mass, which ranges from $1 \ \textrm{M}_\odot$ to $2 \ \textrm{M}_\odot$, in $0.1 \ \textrm{M}_\odot$ increments.}
	\label{fig:hr_diagram_hosts}
\end{figure}
Figure~\ref{fig:hr_diagram_hosts} shows the position of both hosts in a Hertzsprung-Russell diagram.
The figure also includes several MESA Isochrones and Stellar Tracks \citep[MIST;][]{Paxton_2011,Choi_2016,Dotter_2016} evolutionary tracks, with the shown portions corresponding to the subgiant and red-giant branch evolutionary phases.
Each track is color-coded according to its stellar mass, which ranges from $1 \ \textrm{M}_\odot$ to $2 \ \textrm{M}_\odot$, in $0.1 \ \textrm{M}_\odot$ increments.

\subsection{Asteroseismology}
\label{sec:host_astero}

As both targets in this work are notionally solar-like oscillators, we searched for oscillations in their frequency-power spectra.
The power spectra were computed using the corrected light curves produced by the photometric pipeline described in Section~\ref{sec:tess_data}.

Since TESS data for the extended mission (Sector number 27 and above) were observed at a higher cadence of 10 minutes, compared to the 30-min cadence of the nominal mission, we evaluated two different power spectra for each star.
The first power spectrum was created using data from all available sectors from the first year of TESS observations.
The second used all available sectors from the third year (the second year observed the northern ecliptic hemisphere), which resulted in a higher Nyquist frequency due to the faster cadence.

To guide the search for oscillations, and following the procedure of \cite{VanEylen_2016a}, we estimate the location of the frequency of maximum power, $\nu_\textrm{max}$, for each target using the $\nu_\textrm{max} \propto g / \sqrt{T_\textrm{eff}}$ scaling relation \citep{Kjeldsen_1995a}.
The values determined were $\sim 294 \ \mu\textrm{Hz}$ for TOI-4377 and $\sim 363 \ \mu\textrm{Hz}$ for TOI-4551, respectively.
One clear indication from the estimated $\nu_\textrm{max}$ for both targets is that, if there is a detection of the oscillations envelope in the power spectra of the stars, this detection is expected to be visible only on the faster 10-min cadence spectrum, where the Nyquist frequency is $\nu_\textrm{Nyq} \approx 833 \ \mu\textrm{Hz}$ (for 30-min cadence, the value is $\sim 278 \ \mu\textrm{Hz}$).

Unfortunately, visual inspection of all power spectra revealed no visible oscillations envelope, a result that is not entirely surprising given the characteristics of both targets.
Taking into account the results from \cite{Mackereth_2021}, we see that although TOI-4377 has 5 sectors of data available, it has a TESS magnitude of 10.79 and a G magnitude of 11.46, a region of the parameter space where asteroseismic yields are expected to be very low.
On the other hand, TOI-4551, despite being brighter, with a TESS magnitude of 9.03 ($G_\textrm{mag} = 9.64$), it only has 2 sectors of data available, where again the asteroseismic yields are very low.

Finally, we also inspected the asteroseismic catalog from \cite{Hon_2021} to see if there were any asteroseismic estimates for our targets, but neither of them was present in the published data.

\section{Orbital and Planetary Characterization}
\label{sec:planet_characterization}

\subsection{Model adopted}
\label{sec:model_adopted}

To characterize each of the planetary systems we used \textsc{exoplanet}, a \textsc{Python} package for probabilistic modelling of astronomical time series data.
Since both photometric and RV data are available for each of the targets, we simultaneously fit both sets of observations.

For the photometric model of the transit, we considered a stellar quadratic limb darkening, using the analytical model of \cite{Mandel_2002}, with the limb darkening parameters following the parametrization of \cite{Kipping_2013a}.
The parameters included in the model are shown in the left column of Tables~\ref{tab:tic394_joint_params} and \ref{tab:tic204_joint_params}, for TOI-4377 b and TOI-4551 b, respectively, with their respective priors shown in the middle column.
For the period $P$ and time of mid-transit $t_0$ of each planet, the priors were defined as uniform distributions centered roughly around the values determined during the transit search.
The eccentricity $e$, and argument of periastron $\omega$, are sampled uniformly in $\sqrt{e} \cos \omega$ and $\sqrt{e} \sin \omega$ to avoid biases \citep{VanEylen_2016a}.
Both the ratio of radii $R_\textrm{p}/R_\star$, and impact parameter $b$, are sampled uninformatively and uniformly, with the impact parameter having a maximum \textit{a priori} value of $1 \ + \ R_\textrm{p}/R_\star$ to ensure that the planet crosses in front of the star.

Finally, since evolved stars show significant stellar signals on characteristic timescales similar to the transit duration of planets in close orbits \citep{Mathur_2011,Kallinger_2014}, we also take these into account in our photometric model of the transit.
This is done using Gaussian Processes, an approach that has been adopted in the literature \citep{Barclay_2015,Grunblatt_2017} and which has been shown to be capable of characterizing these signals in the time domain \citep{Pereira_2019}.
Following the work of \cite{Pereira_2019}, we consider a stellar signal model with components to capture the mesogranulation and white noise, discarding the component to model the oscillations envelope since it is not detected in the data of either star.
The priors on all parameters from both these components are uniform and adopt identical lower and upper limits to the ones considered in \cite{Pereira_2019}.

For the RV model we consider an RV semi-amplitude $K$ and, for each instrument for which we have available data, a systemic velocity offset $\gamma_\textrm{Instrument}$ and a stellar jitter term $\sigma_\textrm{Instrument}$.
The priors for the systemic velocity offsets are defined as uniform with wide bounds and encompass the median velocity of all data points from each instrument.
As for the semi-amplitude prior, we chose a wide and uniformed uniform prior.
The priors considered for the jitter terms are uniform distributions with an upper limit of the same magnitude as the detected RVs.

\subsection{Parameter estimation}
\label{sec:parameter_estimation}

To sample the posterior probabilities of the model parameters, \textsc{exoplanet} uses \textsc{PyMC3} \citep{Salvatier_2016}, which is a probabilistic framework that implements an Hamiltonian Monte Carlo \citep[HMC;][]{Neal_2011} algorithm, itself a Markov chain Monte Carlo (MCMC) method. 
HMC can compute the derivatives of the posterior distributions, allowing for more efficient sampling.
Additionally, \textsc{PyMC3} also implements a No U-Turn Sampler \citep[NUTS;][]{Hoffman_2011} which improves sampling convergence when there are correlations between parameters.

To ascertain the convergence of the chains, \textsc{exoplanet} provides the Gelman--Rubin statistic, $\hat{R}$ \citep{Gelman_1992}, for each of the sampled variables.
In our models, this value was never larger than 1.01, pointing to a sufficient sampling of all model parameters. 
Additionally, the code also computed bulk and tail effective sample sizes \citep{Gelman_2014} for all sampled properties.
These values allowed us to confirm that all parameter distributions could provide reliable measurements of both bulk properties (median and mean) as well as tail properties (variance and 5\% and 95\% percentiles).

\subsection{Results}
\label{sec:results}


\input{tables/tic394_model_parameters.tex}
\input{tables/tic204_model_parameters.tex}

Tables~\ref{tab:tic394_joint_params} and \ref{tab:tic204_joint_params} show the results obtained in the characterization of TOI-4377 b and TOI-4551 b, respectively.
A definition of the various parameters present in both tables is shown in Table~\ref{tab:symbol_names}.
Some parameters were derived using stellar properties from the planet's host, which are shown in Table~\ref{tab:stellar_properties}.
The tables contain two sets of rows, with the first set containing all parameters present in the model adopted and for which posterior samples were drawn and the second set containing derived parameters obtained using the ones in the previous set.
The middle column contains the prior distributions of all model parameters, whilst the third column contains the posterior distributions, which consist of a median value and a lower and upper bound corresponding to the 16\% (below) and 84\% (above) percentiles.

\begin{figure*}
	\centering
	\includegraphics[width=\textwidth]{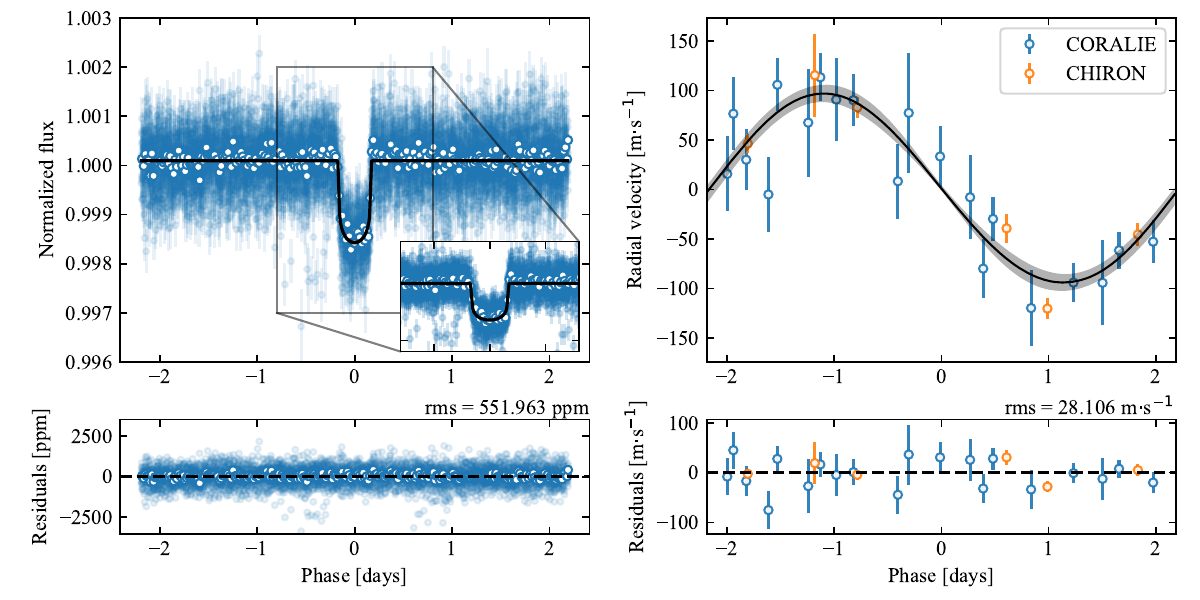}
	\caption{Joint model of photometric and RV observations for TOI-4377 b, phase-folded to the median of the period distribution present in Table~\ref{tab:tic394_joint_params}.
	\textit{Left:} Above, the phase-folded light curve from TESS is shown, with the individual flux observations in blue and the larger white-filled points representing a 10-point temporal binning. 
	The solid black line represents the best fit transit model to the data.
	Below, the residuals between the data and the transit model are shown, with the dashed black line representing the null offset.
	The root-mean-square (rms) deviation between all points and the zero offset is also shown above the plot.
	\textit{Right:} Above, the phase-folded RV observations of TOI-4377 are shown.
	The solid black line is the best fit model for the data, whilst the shaded area represents the 1-$\sigma$ uncertainty.
	The plot below follows the same structure as the one on the left, this time using the residuals between the RV measurements and model shown above.}
	\label{fig:tic394_phasefolded_fit}
\end{figure*}

For TOI-4377 b we find that it orbits its host with a period of $\sim 4.4$ days.
We determine a radius of $1.348 \pm 0.081 \ R_\mathrm{J}$ and a mass of $0.957^{+0.089}_{-0.087} \ M_\mathrm{J}$, which we combine to estimate an average density for the planet of $0.881^{+0.098}_{-0.095} \ \textrm{g} \cdot \textrm{cm}^{-3}$.
Results also point to the orbit having an eccentricity compatible with zero, with the posterior distribution following a half-gaussian with the peak at zero.
To further confirm this, we compute the Lucy-Sweeney test \citep{Lucy_1971}, which determines whether a small $e$ is statistically significant from a circular orbit.
Following \cite{Lucy_1971}, we adopt a $5\%$ level of significance, above which the orbit is assumed to be compatible with zero, and find a value of $\sim 61\%$ for TOI-4377 b.
Figure~\ref{fig:tic394_phasefolded_fit} shows the fit of the model against the phase-folded photometric data, on the left, and the phase-folded RV data, on the right.
The residuals to both fits are shown below their respective plots.
The model was drawn using the median values of the posterior distributions.


\begin{figure*}
	\centering
	\includegraphics[width=\textwidth]{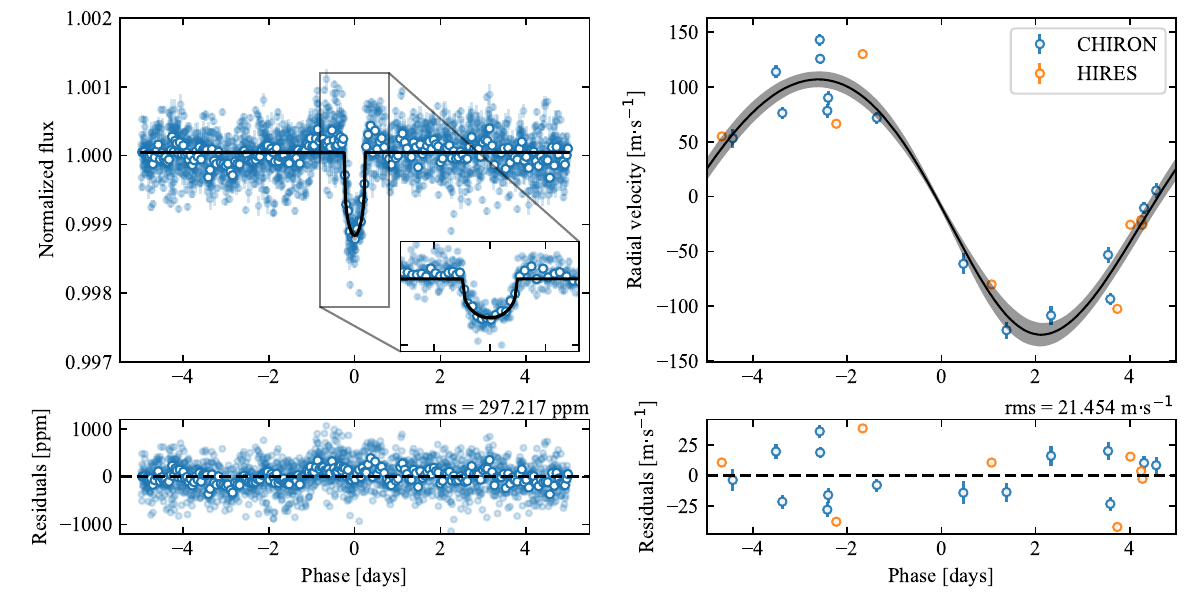}
	\caption{Joint model of photometric and RV observations for TOI-4551 b, phase-folded to the median of the period distribution present in Table~\ref{tab:tic394_joint_params}.
	\textit{Left:} Above, the phase-folded light curve from TESS is shown, with the individual flux observations in blue and the larger white-filled points representing a 10-point temporal binning. 
	The solid black line represents the best fit transit model to the data.
	Below, the residuals between the data and the transit model are shown, with the dashed black line representing the null offset.
	The root-mean-square (rms) deviation between all points and the zero offset is also shown above the plot.
	\textit{Right:} Above, the phase-folded RV observations of TOI-4551 are shown.
	The solid black line is the best fit model for the data, whilst the shaded area represents the 1-$\sigma$ uncertainty.
	The plot below follows the same structure as the one on the left, this time using the residuals between the RV measurements and model shown above.}
	\label{fig:tic204_phasefolded_fit}
\end{figure*}

As for TOI-4551 b, it orbits its evolved host star at a longer period of $\sim 10.0$ days.
For this planet, we find a radius of $1.058^{+0.068}_{-0.062} \ R_\mathrm{J}$ and a mass of $1.49 \pm 0.13 \ M_\textrm{J}$, which we combine to estimate an average density for the planet of $1.74^{+0.19}_{-0.18} \ \textrm{g} \cdot \textrm{cm}^{-3}$.
This planet seems to have a slight eccentricity of $0.10 \pm 0.03$, with a zero-eccentricity solution being over 3-$\sigma$ away.
Again, we compute the Lucy-Sweeney test to this eccentricity estimate, this time obtaining a value of $\sim 0.4\%$, well below the $5\%$ level of significance set to assume a circular orbit.
Figure~\ref{fig:tic204_phasefolded_fit} shows the fit of the model against the phase-folded photometric data, on the left, and the phase-folded RV data, on the right.
The residuals to both fits are shown below their respective plots.
The model was drawn using the median values of the posterior distributions.
Given the low uncertainty in the RV observations, especially for HIRES data, there is a wide gap between the data points and their formal error bars, and the RV model.
However, this gap disappears when adding the RV jitter term estimated for each instrument, in quadrature, to the formal error.
This is not observed for TOI-4377 b as CORALIE's formal errors are much larger.

Additionally, we conducted a search for a second planet in the RV data of TOI-4551.
For that we used the \textsc{RVSearch} search pipeline, presented in \cite{Rosenthal_2021}.
In the pipeline, the False Alarm Probability (FAP) is determined analytically, following the procedure of \cite{Howard_2016}.
From our produced two-planet periodogram, although we find a peak at 33 days, suggesting a possible second, longer period planet, it does not reach a FAP $ < 0.001$.
Thus, we decided that the two-planet fit did not warrant further investigation in this manuscript.

A noticeable feature in the phase-folded light curve of TOI-4551 b (upper left panel of Figure~\ref{fig:tic204_phasefolded_fit}) is the region of phase before the start and after the end of the transit (phases between -1 and 1 days outside of the transit).
Here, the majority of flux measurements is found to be above the continuum line determined in the characterization.
Although we find that this shift is not characterized by any usual models of out of transit variations \citep{Lillo-Box_2014}, we tried to quantify how this shift might affect the planetary radius determined in the fit.
We find that the median offset of these flux measurements is $\sim 150$ ppm, which would lead to a planetary radius of $R_\mathrm{p} = 1.14 R_\mathrm{J}$, slightly more that 1-$\sigma$ above the current uncertainties.
To account for the possibility that this shift is non-stochastic, we revise the presented upper uncertainty in Table~\ref{tab:tic204_joint_params} by adding the increase in radius due to the offset in quadrature to the MCMC derived value.
This leads to a final planetary radius estimation for TOI-4551 b of $R_\mathrm{p} = 1.058^{+0.110}_{-0.062} R_\mathrm{J}$.
Additional TESS observations are needed to reveal whether this feature is significant and related to astrophysical or systematic signals in the data.


\section{Discussion}
\label{sec:discussion}

\subsection{Eccentricity}
\label{sec:eccentricity}

A planet's orbital eccentricity can reveal crucial information on planetary formation by exposing possible events of migration.
\cite{Grunblatt_2018} found preliminary evidence that close-in giant planets orbiting evolved hosts exhibited more eccentric orbits than the same population of planets orbiting main sequence hosts.
Later, \cite{Grunblatt_2023} studied the population of transiting planets that orbit evolved hosts and observed that they seem to follow a linear trend in the logperiod-eccentricity plane (see Figure~6 of \cite{Grunblatt_2023}).
He noted that planet-planet scattering events could be exciting planetary orbital eccentricities at longer periods whilst planet-star interactions would lead to tidal circularization at smaller periods.

Looking at the targets in this work, the eccentricity of TOI-4377 b, which is compatible with zero (see note in Table~\ref{tab:tic394_joint_params}) and hint of low eccentricity of TOI-4551 b match the observed period-eccentricity trend found by \cite{Grunblatt_2023}.
Moderate eccentricities of hot/warm Jupiters are expected for planets that are experiencing tidal migration \citep{Villaver_2014}. 
The eccentricity of TOI-4551 may indicate a dynamic past for this system, which could be supported by the detection of additional companions.
The unusually high amount of RV scatter for this star as seen in the Keck/HIRES measurements may also be indicative of an unseen companion.

\subsection{Radius Inflation}
\label{sec:inflation}

\begin{figure*}
	\centering
	\includegraphics[width=\columnwidth]{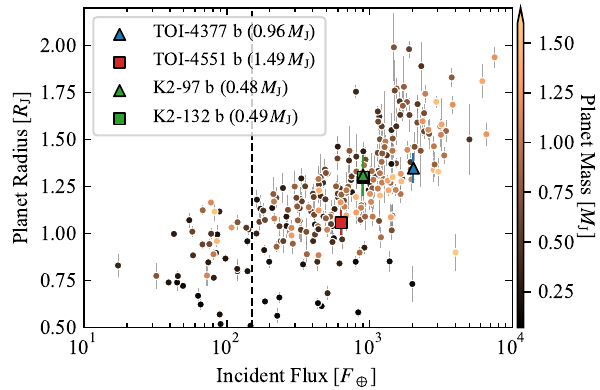}
	\includegraphics[width=\columnwidth]{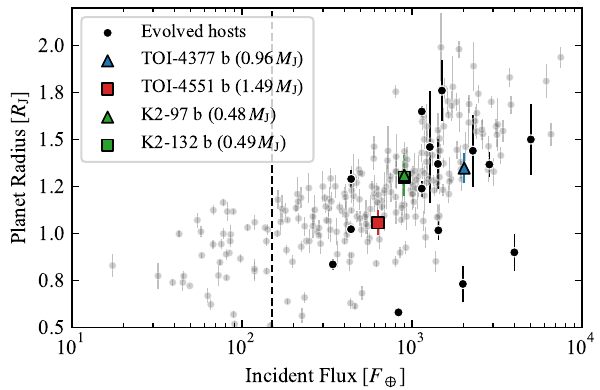}
	\caption{\textit{Left:} Incident flux received by a planet against its radius, for close-in (<20 days) giant ($>$$0.5 \ R_\mathrm{J}$) planets.
	The color-coding denotes the planetary mass.
	TOI-4377 b, TOI-4551 b, K2-97 b and K2-132 b are identified and their masses are shown in the label.
	The vertical black dashed line shows the empirical incident flux cutoff for planetary radius inflation \citep{Demory_2011,Miller_2011}.
	Data for confirmed planets was obtained from the NASA Exoplanet Archive.
	\textit{Right:} Same dataset as the left panel, this time highlighting evolved systems. 
	Grey circles represent planets orbiting dwarf stars whilst the larger black circles show planets with evolved hosts ($R_\star > 2.0 \, \mathrm{R}_\odot$, $T_\mathrm{eff} < 6000 \,$K).}
	\label{fig:incident_flux}
\end{figure*}
%

Planetary radius inflation has been observed in both main sequence and evolved systems, suggesting that it occurs independently of the system's evolutionary state.
The left panel of figure~\ref{fig:incident_flux} places both TOI-4377 b and TOI-4551 b in context with other known short-orbit (<20 days) Jupiter-sized (>0.5 $R_\mathrm{J}$) planets, showing planetary radius against the incident flux from its host, with the planetary mass shown through color-coding.
From the figure, the first noticeable feature is the clear correlation between planet radius and the incident flux from the planet's host star, which is well-documented in the literature \citep{Laughlin_2011,Lopez_2016,Thorngren_2018}.
Both TOI-4377 b and TOI-4551 b fit well within the existing population and the observed curve, with TOI-4377 b having an incident flux of $2027 \pm 160 \ F_\oplus$ and TOI-4551 b of $631 \pm 67 \ F_\oplus$.

The literature also notes a relation between planet radius and mass in the context of inflation \citep{Thorngren_2018,Thorngren_2021}, with higher mass planets having smaller radii when subjected to the same stellar incident flux.
\cite{Thorngren_2021} point to the higher gravity of higher mass planets as responsible for offsetting some of the effects of the planet's interior heating (and subsequent radius inflation).
This correlation can also be observed in the left panel of Figure~\ref{fig:incident_flux}, where darker points (lower mass) appear higher vertically (higher radii).

Turning to the right panel of Figure~\ref{fig:incident_flux}, the focus is now on the population of known transiting hot Jupiters orbiting evolved hosts ($R_\star > 2.0 \, \mathrm{R}_\odot$, $T_\mathrm{eff} < 6000 \,$K; shown as black circles), a population which includes TOI-4377 b and TOI-4551 b (shown as the blue triangle and red square, respectively).
In general, we observe that these planets follow the same trend with stellar incident flux mentioned above, with all planets having incident fluxes above the black dashed vertical line which shows the empirical incident flux cutoff for planetary radius inflation \citep{Demory_2011,Miller_2011}.

\cite{Lopez_2016} postulated that evolved systems with incident fluxes between $500 \ F_\oplus - 1500 \ F_\oplus$ would be very valuable in quantifying the efficiency of planetary radius re-inflation.
\cite{Grunblatt_2017} followed by estimating the stellar irradiation heating efficiency through the characterization of two \textit{K2} planets, K2-97 b and K2-132 b (shown in the figures as the overlapping green triangle and square, respectively).
Compared to those \textit{K2} planets, which had similar masses, radii, orbits and stellar hosts, our two planets have different masses, radii and stellar incident fluxes, with only the lower mass one having a clearly inflated radius \citep[$R_\mathrm{p} > 1.2 \, R_\mathrm{J}$][]{Fortney_2007}.

For this reason, we sought to characterize the efficiency with which the stellar incident flux is deposited in the planetary interior of our planets.
To do so, models of planetary interiors were produced for both planets, following the approach of \cite{Thorngren_2018}.
These planetary interior structure models include an additional heating power source that characterizes the energy from the stellar irradiation that is deposited in the planet's interior and is identified by a heating efficiency quantity, $\epsilon(F)$, which represents a fraction of the stellar incident flux, $F$.

The posterior distributions for the parameters of the interior structure models for each star are shown in Figures~\ref{fig:toi4377_inflation_corner} and \ref{fig:toi4551_inflation_corner} for TOI-4377 b and TOI-4551 b, respectively.
From the results, we estimate a heating efficiency of $1.91 \pm 0.48\%$ and $2.19 \pm 0.45\%$ for TOI-4377 b and TOI-4551 b respectively.
Additionally, the interior models also fit for the heavy-element fraction of the planets, and we measure values of $0.26 \pm 0.04$ for TOI-4377 b and $0.27 \pm 0.05$ for TOI-4551 b, which are relatively high values but not completely unexpected for planets of this mass (see Figure~10 of \citealt{Thorngren_2016}).


Driven by the typical radii and incident fluxes of our planets, the interior models yield heating efficiencies and planetary bulk metallicities in line with the known population of hot Jupiters, including hot Jupiters orbiting main sequence hosts.
This suggests that the current radii of both our planets reflect the current stellar incident flux to which they are exposed, which has increased as the stellar host entered the post-main-sequence.
This in turn suggests that reinflation is happening at a fast enough rate that our planet's radii have inflated alongside the increase in their hosts' brightness.
Our results are consistent with the ones of \cite{Thorngren_2021}, which find the same conclusion that some degree of reinflation is present for hot Jupiters as their host star brightens throughout the main sequence.

Furthermore, our heating efficiency values are also in good agreement with the heating efficiency inferred for re-inflated hot Jupiters transiting evolved stars from \textit{K2}, K2-97 b and K2- 132 b \citep{Grunblatt_2017}.
Comparing our planets to these two, we can see some the impact of planetary mass and incident flux on the degree of radius inflation \citep{Thorngren_2018}.
TOI-4377 b is subjected to a higher incident flux than the \textit{K2} planets but is two times more massive and the resulting radii between all planets is roughly the same.
As for TOI-4551 b, it has the largest mass and the lowest incident flux of all planets in the comparison, and does not appear to have an inflated radius (looking at its radius only) despite all planets having similar heating efficiencies.

Still, there are some caveats to the interior structure models presented here.
The interior models attempt to estimate both the heating efficiency and bulk metallicity of the planets from the planetary radius alone.
This degeneracy is mitigated by defining populational priors on the planet bulk metallicity \citep{Thorngren_2016} and the heating efficiency \citep{Thorngren_2018}.
Nevertheless, if a planet is smaller than the typical, the model can't tell whether that's because of an above-average metallicity or a below average heating efficiency, although the Bayesian model accounts for this problem in a statistically rigorous way.

\subsection{Tidal Evolution}
\label{sec:tidal_evolution}

Tidal interactions between a planet and its host star can cause planets to spiral inwards and may eventually lead to tidal destruction \citep{Villaver_2009,Schlaufman_2013}.
One possible observation of this phenomenon is the dearth of close-in giant planets orbiting evolved stars in RV surveys \citep{Johnson_2010a,Jones_2016}, which hinted at possible differences in the populations of these planets orbiting evolved and main sequence hosts.
The larger radii of these evolved hosts, when compared to their main sequence counterparts, would correspond to an increase in the strength of the tidal interactions (see Equation~\ref{eq:tidal_decay}), leading to quicker timescales in the spiralling and destruction and consequently to a reduced number of the shortest period hot Jupiters \citep{Hamer_2019}.
Despite this, hot Jupiter occurrence rates appear similar between main sequence and evolved hosts \citep{Grunblatt_2019,Temmink_2023}.


%
Still, it is interesting to try and estimate the timescales of tidal decay and eventual destruction expected for both TOI-4377 b and TOI-4551 b, which can give an indication of the expected period variation of the planet and of the tidal dissipation strength in the system.
To do so, we follow Equation~(3) from \cite{VanEylen_2016a} (adapted from Eq.~(11) of \cite{Schlaufman_2013}) to estimate the timescale of orbital decay, $t$, with:
\begin{equation}
	t = 10 \, \textrm{Gyr} \frac{Q_\star / k_\star}{10^6} %
	\left( \frac{M_\star}{\textrm{M}_\odot} \right)^{1/2} %
	\left( \frac{M_\textrm{p}}{M_\textrm{J}} \right)^{-1} %
	\left( \frac{R_\star}{\textrm{R}_\odot} \right)^{-5} %
	\left( \frac{a}{0.06 \ \textrm{au}} \right)^{13/2} .
	\label{eq:tidal_decay}
\end{equation}

The quantities $Q_\star$ and $k_\star$ represent the stellar tidal quality factor and the second-order stellar tidal Love number respectively.
This same quantity is often represented in the literature through the modified tidal quality factor, $Q'_\star = 3 \, Q_\star / 2 \, k_\star$ \citep{Yee_2020}.
Considering a nominal value of $Q'_\star = 10^6$, we estimate that, for TOI-4377 b, the timescale for tidal decay is $\sim$$1.8 \times 10^7$ years.
For TOI-4551 b, which has a longer orbit, the value is $\sim$$4.1 \times 10^8$ years.
Both measurements are dependent on the choice of the tidal quality factor.

Some measurements of other systems that provide constraints to this modified tidal quality factor already exist within the literature.
\cite{Maciejewski_2016} used transit data of WASP-12 b, which orbits a Sun-like star, to estimate a value of $Q'_\star = 2.5 \times 10^5$, with more recent works finding similar values (\citealt{Patra_2017} determined a value of $Q'_\star = 2 \times 10^5$, and \citealt{Yee_2020} a value of $Q'_\star = 1.75 \times 10^5$).
The few existing estimates for this quantity place it above $10^5$ \citep[e.g.][]{Maciejewski_2018,Patra_2020}.

For evolved stars, \cite{Schlaufman_2013} estimated that values of the modified tidal decay factor should be on the order of $10^2 - 10^3$, but measurements seem to point to similar values to the ones found for planets around main sequence stars.
\cite{Chontos_2019} found that $Q'_\star = 1.2 \times 10^5$ for Kepler-1658 b, whilst \cite{Grunblatt_2022} estimated a lower limit of $Q'_\star > 2 \times 10^4$ for TOI-2337 b.
Considering this later, lower value when estimating the timescales of our planets, we find that the timescales for planetary tidal engulfment are $\sim$$3.6 \times 10^5$ years and $\sim$$8.3 \times 10^6$ years for TOI-4377 b and TOI-4551 b, respectively.
For TOI-4377 b in particular, the shorter period planet of the two, this would correspond to a yearly decrease in its measured period of $\sim$1 second.

Given the short period and multiple sectors available for TOI-4377 b, we also attempted to measure an orbital decay rate from the individual transit times, but we find no evidence of a shift larger than the statistical scatter in the transit times themselves ($\sim 0.01$ days).
Currently, given that the transit times for this target are condensed in $\sim2-3$ month time spans, two years apart, attempts to measure a lower limit of $Q'_\star$ have not been successful.
This is due to a degeneracy in the fit as the modelled quadratic curve does not have any data points near its inflection point to constrain it, allowing the curve of the function to essentially be unbounded.
As TESS data from future extended missions becomes available for this target, this degeneracy will be broken, which should enable the calculation of a lower limit for the value of $Q'_\star$, so that it can be compared with others in the literature.
Additionally, the next TESS extended mission will observe FFIs at a 200-sec cadence, further improving the precision in the determination of the individual transit times.


\section{Conclusions}
\label{sec:conclusions}

We report the identification and confirmation of two close-in giant planets orbiting red-giant stars, TOI-4377 b and TOI-4551 b.
The planets were found during a search for transits around bright, low-luminosity red-giant branch stars observed by TESS in the southern ecliptic hemisphere (during the first year of the nominal mission).
Both targets were followed-up with RV observations which confirmed their planetary nature and improved the characterization of the system, despite the lack of asteroseismic signal detections in the data.

	TOI-4377 b is a hot Jupiter with a radius of $1.35 \ R_\textrm{J}$ and a mass of $0.96 \ M_\textrm{J}$.
	It orbits its host, a red-giant star ($3.52 \ \textrm{R}_\odot$, $1.36 \ \textrm{M}_\odot$), every $4.38$ days with an orbit with an eccentricity compatible with zero (see note in Table~\ref{tab:tic394_joint_params}).
	Its clearly inflated radius and orbit around an evolved star (which corresponds to a higher irradiation of the planet, when compared to a dwarf host) make this a good candidate for studies on planetary radius re-inflation, and we produce planetary interior structure models of the planet and estimate an inflation heating efficiency of $1.91 \pm 0.48\%$.
	These results point to stellar irradiation being a capable mechanism to explain the inflated radii of hot Jupiters, and that this reinflation is happening at a pace faster than post-main sequence evolution.
	Using a tentative lower limit value for the modified tidal decay factor we estimate a tidal decay timescale for the planet of $\sim$$3.6 \times 10^5$ years, which corresponds to a yearly change in the period of the planet of $\sim 1$ second.
	Given the short period of this planet, we attempt to determine a lower limit to the tidal decay factor from the individual transit times, but conclude that due to degeneracies in the dataset, no valuable estimate can be produced currently.

	TOI-4551 b is a massive hot Jupiter with with a radius of $1.06 \ R_\textrm{J}$ and a mass of $1.49 \ M_\textrm{J}$.
	It orbits its host, a red-giant star ($3.55 \ \textrm{R}_\odot$, $1.31 \ \textrm{M}_\odot$), every $9.95$ days with an orbit with eccentricity of $0.10$.
	From planetary interior models we find a radius inflation efficiency of $2.19 \pm 0.45\%$, suggesting that radius inflation is also taking place in the planet, despite its modest radius.
	When compared to the \textit{K2} giant planets K2-97 b and K2-132 b, which also orbit evolved stars \citep{Grunblatt_2017}, we see how both its larger mass and lower incident flux influence the size of the planet, since despite having similar heating efficiencies to both \textit{K2} planets, its radius is smaller and even seemingly non-inflated ($R_\mathrm{p} < 1.2 \, R_\mathrm{J}$).
	Using the same assumptions on the tidal decay factor as for TOI-4377 b, we estimate a tidal decay timescale for this planet of $\sim$$8.3 \times 10^6$ years.

All in all, the two newly confirmed planets enrich the existing population of known hot Jupiters orbiting evolved hosts.
They should prove particularly relevant to demographic studies of short-period giant planets around red-giant stars, where only a handful of planets are known.
At the same time, the systems' unique properties also make them interesting targets to further decipher the mechanisms behind planetary radius inflation and might possibly lead to measurements of orbital decay through tidal interactions between planet and host.

\section*{Acknowledgements}

This work was supported by FCT - Fundação para a Ciência e a Tecnologia through national funds and by FEDER through COMPETE2020 - Programa Operacional Competitividade e Internacionalização by these grants: UIDB/04434/2020; UIDP/04434/2020.
S.K.G. acknowledges support from the National Aeronautics and Space Administration under grants 80NSSC19K0593 and 80NSSC21K0781 issued through the TESS Guest Investigator Program.
TLC is supported by FCT in the form of a work contract (CEECIND/00476/2018).
MSC acknowledges the support from the FCT through a work contract (CEECIND/02619/2017).
Co-funded by the European Union (ERC, FIERCE, 101052347). 
Views and opinions expressed are however those of the author(s) only and do not necessarily reflect those of the European Union or the European Research Council.
Neither the European Union nor the granting authority can be held responsible for them. 
ML acknowledges support of the Swiss National Science Foundation under grant number PCEFP2\_194576.
SQ acknowledges support from the TESS GI Program under award 80NSSC21K1056 and from the TESS mission via subaward s3449 from MIT.
We thank the Swiss National Science Foundation (SNSF) and the Geneva University for their continuous support to our planet low-mass companion search programs. 
This work was been in particular carried out within the framework of the Swiss National Centre for Competence in Research (NCCR) $PlanetS$ supported by the Swiss National Science Foundation (SNSF) under grants 51NF40$\_$182901 and 51NF40$\_$205606. 
ML and BA acknowledge support of the Swiss National Science Foundation under grant number PCEFP2$\_$194576. 
This publication makes use of The Data $\&$ Analysis Center for Exoplanets (DACE), which is a facility based at the University of Geneva (CH) dedicated to extrasolar planet data visualization, exchange, and analysis. 
DACE is a platform of NCCR $PlanetS$ and is available at \url{https://dace.unige.ch}.
Some of the observations in this paper made use of the High-Resolution Imaging instrument Zorro and were obtained under Gemini LLP Proposal Number: GN/S-2021A-LP-105. 
Zorro was funded by the NASA Exoplanet Exploration Program and built at the NASA Ames Research Center by Steve B. Howell, Nic Scott, Elliott P. Horch, and Emmett Quigley. 
Zorro was mounted on the Gemini South telescope of the international Gemini Observatory, a program of NSF's OIR Lab, which is managed by the Association of Universities for Research in Astronomy (AURA) under a cooperative agreement with the National Science Foundation. 
On behalf of the Gemini partnership: the National Science Foundation (United States), National Research Council (Canada), Agencia Nacional de Investigación y Desarrollo (Chile), Ministerio de Ciencia, Tecnología e Innovación (Argentina), Ministério da Ciência, Tecnologia, Inovações e Comunicações (Brazil), and Korea Astronomy and Space Science Institute (Republic of Korea).
This paper made use of data collected by the TESS mission and are publicly available from the Mikulski Archive for Space Telescopes (MAST) operated by the Space Telescope Science Institute (STScI). 
Funding for the TESS mission is provided by NASA's Science Mission Directorate.
We acknowledge the use of public TESS data from pipelines at the TESS Science Office and at the TESS Science Processing Operations Center. 
Resources supporting this work were provided by the NASA High-End Computing (HEC) Program through the NASA Advanced Supercomputing (NAS) Division at Ames Research Center for the production of the SPOC data products.
D.H. acknowledges support from the Alfred P. Sloan Foundation, the National Aeronautics and Space Administration (80NSSC21K0652), and the Australian Research Council (FT200100871).

\textit{Software:} This research made use of open-source software, namely: \textsf{exoplanet} \citep{Foreman-Mackey_2021,Foreman-Mackey_2021b} and its dependencies \citep{Foreman-Mackey_2017,Foreman-Mackey_2018,Agol_2020,Kumar_2019,AstropyCollaboration_2013,AstropyCollaboration_2018,Kipping_2013a,Luger_2019,Salvatier_2016,TheTheanoDevelopmentTeam_2016}, \textsc{Tesscut} \citep{Brasseur_2019}, \textsc{astropy} \citep{AstropyCollaboration_2013,AstropyCollaboration_2018}, \textsc{eleanor} \citep{Feinstein_2019}, \textsc{transitleastsquares} \citep{Hippke_2019} and \textsc{VESPA} \citep{Morton_2012,Morton_2015b}.


\section*{Data Availability}
 
The TESS FFI data used in this article were accessed from MAST portal (Barbara A. Mikulski Archive for Space Telescopes) at \url{https://mast.stsci.edu/portal/Mashup/Clients/Mast/Portal.html}. 
As mentioned in Section~\ref{sec:tess_data}, a full description of the aperture photometry and planet search pipeline used in this work, including a description and links to all open-source software used in it is described in detail in \cite{Pereira_2022}.
Imaging data from \textit{Gemini} are accessible through the ExoFOP portal at \url{https://exofop.ipac.caltech.edu/tess/target.php?id=394918211} for TOI-4377 and at \url{https://exofop.ipac.caltech.edu/tess/target.php?id=204650483} for TOI-4551, respectively.
The derived data generated for this research and corresponding code to produce the figures will be shared upon reasonable request from the corresponding author. 


\bibliographystyle{mnras}
\bibliography{main}


\appendix

\section{Radial-velocity measurements}

\input{tables/tic394_rv_obs.tex}
\input{tables/tic204_rv_obs.tex}

\section{Symbol names}

\input{tables/symbol_names.tex}

\section{Inflation model posterior distributions}

\begin{figure}
	\centering
	\includegraphics[width=\columnwidth]{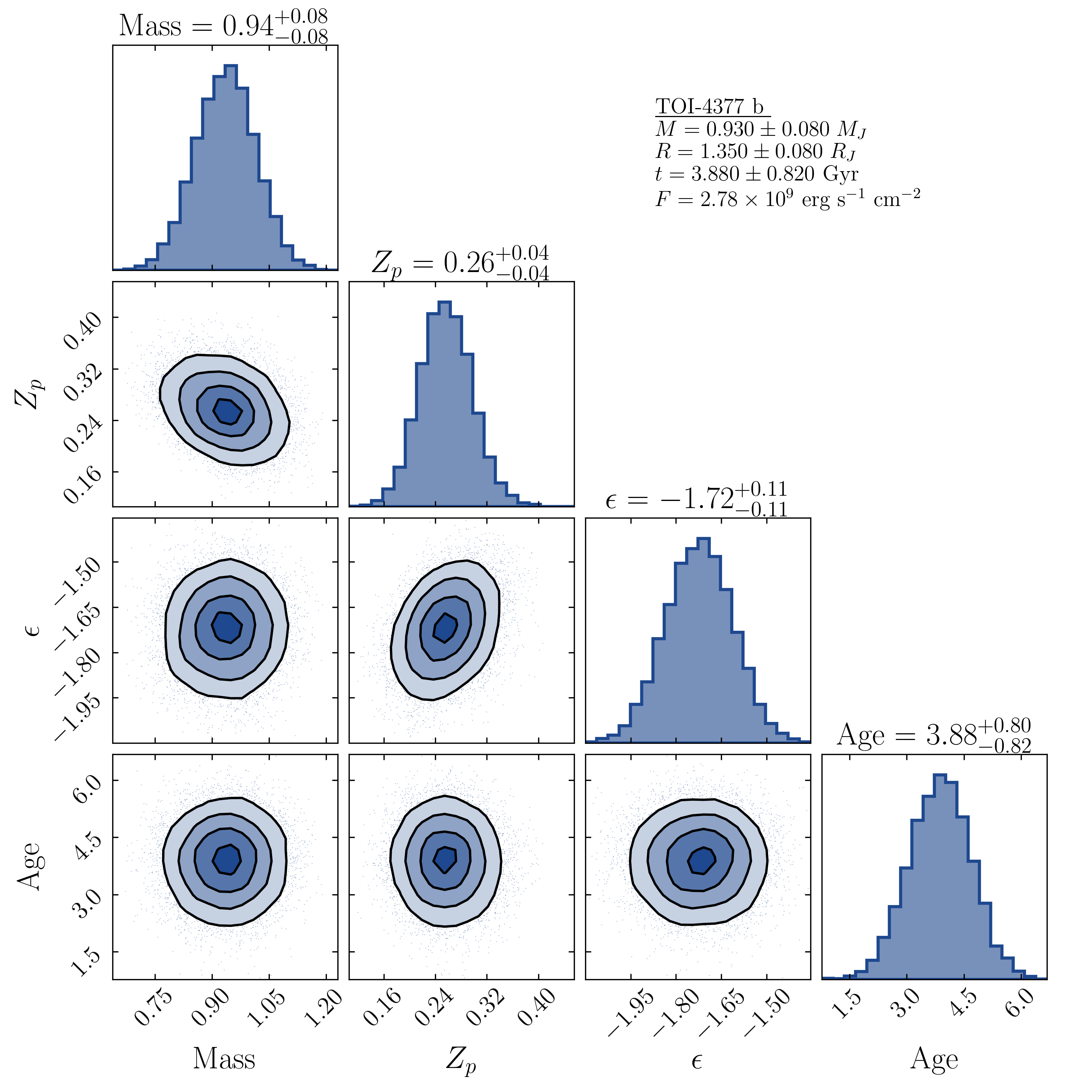}
	\caption{Posterior distributions for the parameters of the inflation model of TOI-4377 b.
	Top to bottom, the parameters are the mass of the planet, observed fraction of metals in the planet ($Z_\mathrm{p}$), log of the inflation efficiency ($\epsilon$) and age of the system.}
	\label{fig:toi4377_inflation_corner}
\end{figure}
\begin{figure}
	\centering
	\includegraphics[width=\columnwidth]{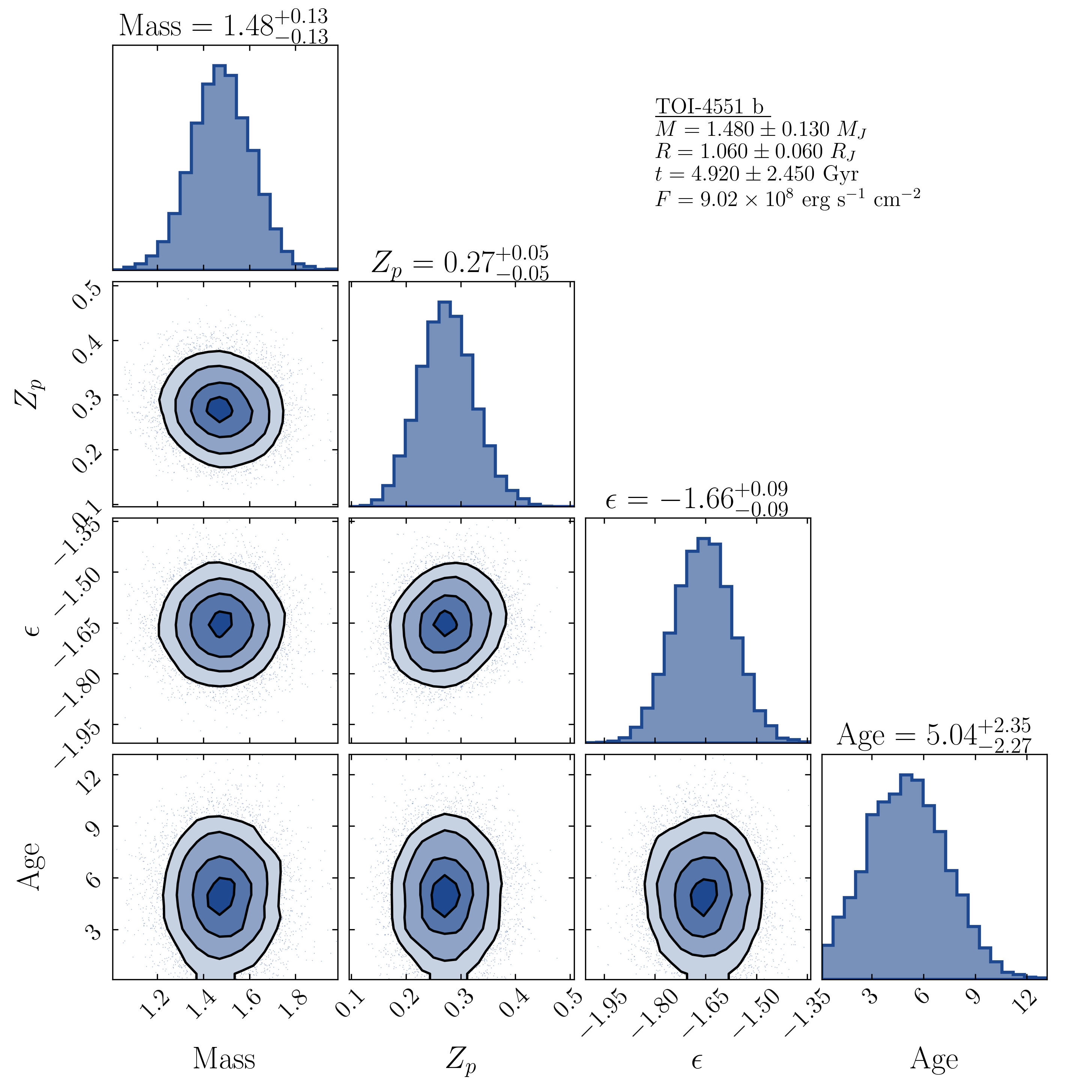}
	\caption{Posterior distributions for the parameters of the inflation model of TOI-4551 b.
	The parameters are the same as in Figure~\ref{fig:toi4377_inflation_corner}.}
	\label{fig:toi4551_inflation_corner}
\end{figure}


\bsp	
\label{lastpage}
\end{document}

%% file: tables/stellar_properties.tex
\begin{table}
    \centering
    \setlength{\tabcolsep}{6pt}
    \renewcommand{\arraystretch}{1.1}
    \caption{Stellar parameters for TOI-4377 and TOI-4551}
    \begin{tabular}{lcc}
        \toprule
        \toprule
        Parameter           & TOI-4377              & TOI-4551              \\
        \midrule
        \multicolumn{3}{c}{Basic properties}                                \\
        \midrule
        TIC                 & 394918211             & 204650483             \\
        2MASS               & J10594390-8213171     & J12195494-2549399     \\
        Gaia DR2            & 5197568146570178176   & 3476605537386300032   \\
        R.A.                & 10:59:43.79           & 12:19:54.9            \\
        Decl.               & $-$82:13:16.86        & $-$25:49:40.27        \\
        Magnitude (TESS)    & 10.79                 & 9.03                  \\
        Magnitude ($V$)     & 11.71\textsuperscript{\textasteriskcentered} & 9.97\textsuperscript{\textdagger} \\
        \midrule
        \multicolumn{3}{c}{\textit{Gaia} DR3 Parallax}                      \\
        \midrule
        $\pi$ (mas)                       & $2.179 \pm 0.012$         & $4.594 \pm 0.017$         \\
        $L_\star$ ($\mathrm{L}_\odot$)    & $6.64 \pm 0.48$           & $6.35 \pm 0.44$           \\
        \midrule
        \multicolumn{3}{c}{Spectroscopy}                                    \\
        \midrule
        $T_\mathrm{eff}$ (K)              & $4974 \pm 57$             & $4896 \pm 110$            \\
        $\log g$ (dex)                    & $3.39 \pm 0.16$           & $3.48 \pm 0.10$           \\
        $[$Fe/H$]$ (dex)                  & $0.205 \pm 0.038$         & $0.23 \pm 0.06$           \\
        $v \sin i$ (km$\cdot$s$^{-1}$)    & $1.260 \pm 0.069$         & $2.84 \pm 1.00$           \\
        $M_\star$ ($\mathrm{M}_\odot$)    & $1.36^{+0.08}_{-0.07}$    & $1.31^{+0.07}_{-0.18}$    \\
        $R_\star$ ($\mathrm{R}_\odot$)    & $3.52^{+0.15}_{-0.14}$    & $3.55^{+0.18}_{-0.15}$    \\
        $\rho_\star$ (g$\cdot$cm$^{-3}$)  & $0.031^{+0.005}_{-0.004}$ & $0.028^{+0.006}_{-0.006}$ \\
        $t$ (Gyr)                         & $3.88^{+0.82}_{-0.54}$    & $4.92^{+2.45}_{-1.33}$    \\
        \bottomrule
        \multicolumn{3}{l}{
            \textsuperscript{\textasteriskcentered} Additional magnitudes in \href{https://exofop.ipac.caltech.edu/tess/target.php?id=394918211}{ExoFOP TIC 394918211}
        } \\
        \multicolumn{3}{l}{
            \textsuperscript{\textdagger} Additional magnitudes in \href{https://exofop.ipac.caltech.edu/tess/target.php?id=204650483}{ExoFOP TIC 204650483}
        } \\
    \end{tabular}
    \label{tab:stellar_properties}
\end{table}

%% file: tables/tic394_model_parameters.tex
\begin{table}
    \centering
    \setlength{\tabcolsep}{6pt}
    \renewcommand{\arraystretch}{1.1}
    \caption{Model and Derived Parameters for TOI-4377 b}
    \begin{tabular}{lll}
        \toprule
        \toprule
        Parameter                                          & Prior                                      & Posterior                            \\
        \midrule
        \multicolumn{3}{c}{Model Parameters}                                                                                                   \\
        \midrule
        $P$ (days)                                         & $\mathcal{U}$(4.200000, 4.500000)          & 4.378081$\pm$0.000019                \\ 
        $t_0$ (BTJD)                                       & $\mathcal{U}$(1601.0000, 1603.0000)        & 1602.1245$\pm$0.0023                 \\ 
        $\sqrt{e} \cos \omega$                             & $\mathcal{U}$(-1.00, 1.00)                 & 0.06$^{+0.12}_{-0.13}$               \\ 
        $\sqrt{e} \sin \omega$                             & $\mathcal{U}$(-1.00, 1.00)                 & -0.08$^{+0.21}_{-0.19}$              \\ 
        $b$                                                & $\mathcal{U}$(0.000, 1.000 + $R_p/R_\star$) & 0.605$^{+0.081}_{-0.100}$            \\ 
        $u_1$                                              & $\mathcal{U}$(-1.00, 1.00)                 & 0.27$^{+0.28}_{-0.19}$               \\ 
        $u_2$                                              & $\mathcal{U}$(-1.00, 1.00)                 & 0.30$^{+0.30}_{-0.38}$               \\ 
        $R_\mathrm{p} / R_\star$                           & $\mathcal{U}$(0.00100, 0.10000)            & 0.03920$^{+0.00100}_{-0.00097}$      \\ 
        $\log\sigma_\mathrm{TESS}$                         & $\mathcal{U}$(-12.000, -2.000)             & -7.615$^{+0.011}_{-0.010}$           \\ 
        $a_\mathrm{gran}$ (ppm)                            & $\mathcal{U}$(20, 400)                     & 259$^{+14}_{-13}$                    \\ 
        $b_\mathrm{gran}$ ($\mu$Hz)                        & $\mathcal{U}$(1.00, 100.00)                & 10.26$^{+1.10}_{-0.97}$              \\ 
        $K$ (m$\cdot$s$^{-1}$)                             & $\mathcal{U}$(-250.0, 250.0)               & 96.2$^{+8.1}_{-8.0}$                 \\ 
        $\gamma_\mathrm{CORALIE}$ (m$\cdot$s$^{-1}$)       & $\mathcal{U}$(12000.0, 13000.0)            & 12554.4$^{+7.5}_{-7.6}$              \\ 
        $\gamma_\mathrm{CHIRON}$ (m$\cdot$s$^{-1}$)        & $\mathcal{U}$(10500, 11500)                & 11048$^{+12}_{-11}$                  \\ 
        $\sigma_\mathrm{CORALIE}$ (m$\cdot$s$^{-1}$)       & $\mathcal{U}$(0.1, 100.0)                  & 9.9$^{+10.0}_{-6.8}$                 \\ 
        $\sigma_\mathrm{CHIRON}$ (m$\cdot$s$^{-1}$)        & $\mathcal{U}$(0, 100)                      & 23$^{+17}_{-11}$                     \\ 
        \midrule
        \multicolumn{3}{c}{Derived Properties}                                                                                                 \\
        \midrule
        $e$                                                &                                            & 0.045$^{+0.053}_{-0.032}$\textasteriskcentered \\ 
        $\omega$ (deg)                                     &                                            & -0.73$^{+2.00}_{-0.98}$              \\ 
        $a$ (AU)                                           &                                            & 0.0579$^{+0.0010}_{-0.0011}$         \\ 
        $i$ (deg)                                          &                                            & 80.3$^{+1.7}_{-1.4}$                 \\ 
        $R_\mathrm{p}$ ($R_\mathrm{J}$)                  &                                            & 1.348$\pm$0.081                      \\ 
        $M_\mathrm{p}$ ($M_\mathrm{J}$)                  &                                            & 0.957$^{+0.089}_{-0.087}$            \\ 
        $\rho_\mathrm{p}$ (g$\cdot$cm$^{-3}$)              &                                            & 0.881$^{+0.098}_{-0.095}$            \\ 
        \bottomrule
        \bottomrule
        \multicolumn{3}{l}{
            \textsuperscript{\textasteriskcentered} The posterior samples show an half-normal distribution centered on 0.
        } \\
    \end{tabular}
    \label{tab:tic394_joint_params}
\end{table}

%% file: tables/tic204_model_parameters.tex
\begin{table}
    \centering
    \setlength{\tabcolsep}{6pt}
    \renewcommand{\arraystretch}{1.1}
    \caption{Model and Derived Parameters for TOI-4551 b}
    \begin{tabular}{lll}
        \toprule
        \toprule
        Parameter                                          & Prior                                      & Posterior                            \\
        \midrule
        \multicolumn{3}{c}{Model Parameters}                                                                                                   \\
        \midrule
        $P$ (days)                                         & $\mathcal{U}$(9.800000, 10.100000)         & 9.955810$^{+0.000076}_{-0.000078}$   \\ 
        $t_0$ (BTJD)                                       & $\mathcal{U}$(1574.5000, 1576.5000)        & 1575.6597$^{+0.0051}_{-0.0044}$      \\ 
        $\sqrt{e} \cos \omega$                             & $\mathcal{U}$(-1.000, 1.000)               & -0.263$^{+0.078}_{-0.055}$           \\ 
        $\sqrt{e} \sin \omega$                             & $\mathcal{U}$(-1.00, 1.00)                 & 0.12$^{+0.14}_{-0.19}$               \\ 
        $b$                                                & $\mathcal{U}$(0.00, 1.00 + $R_p/R_\star$)  & 0.31$^{+0.17}_{-0.19}$               \\ 
        $u_1$                                              & $\mathcal{U}$(-1.00, 1.00)                 & 0.76$^{+0.25}_{-0.30}$               \\ 
        $u_2$                                              & $\mathcal{U}$(-1.00, 1.00)                 & 0.01$^{+0.38}_{-0.34}$               \\ 
        $R_\mathrm{p} / R_\star$                           & $\mathcal{U}$(0.0010, 0.1000)              & 0.0308$^{+0.0011}_{-0.0010}$         \\ 
        $\log\sigma_\mathrm{TESS}$                         & $\mathcal{U}$(-12.000, -2.000)             & -8.268$^{+0.020}_{-0.019}$           \\ 
        $a_\mathrm{gran}$ (ppm)                            & $\mathcal{U}$(20, 400)                     & 152$\pm$11                           \\ 
        $b_\mathrm{gran}$ ($\mu$Hz)                        & $\mathcal{U}$(1.0, 100.0)                  & 17.5$^{+4.0}_{-3.5}$                 \\ 
        $K$ (m$\cdot$s$^{-1}$)                             & $\mathcal{U}$(-250.0, 250.0)               & 117.3$^{+8.5}_{-8.2}$                \\ 
        $\gamma_\mathrm{CHIRON}$ (m$\cdot$s$^{-1}$)        & $\mathcal{U}$(41000.0, 42000.0)            & 41292.7$^{+6.0}_{-6.1}$              \\ 
        $\gamma_\mathrm{HIRES}$ (m$\cdot$s$^{-1}$)         & $\mathcal{U}$(-500, 500)                   & 0$^{+12}_{-13}$                      \\ 
        $\sigma_\mathrm{CHIRON}$ (m$\cdot$s$^{-1}$)        & $\mathcal{U}$(0.1, 100.0)                  & 22.0$^{+6.2}_{-4.4}$                 \\ 
        $\sigma_\mathrm{HIRES}$ (m$\cdot$s$^{-1}$)         & $\mathcal{U}$(0.1, 100.0)                  & 32.7$^{+13.0}_{-7.9}$                \\ 
        \midrule
        \multicolumn{3}{c}{Derived Properties}                                                                                                 \\
        \midrule
        $e$                                                &                                            & 0.101$^{+0.034}_{-0.033}$            \\ 
        $\omega$ (deg)                                     &                                            & 2.37$^{+0.48}_{-5.10}$               \\ 
        $a$ (AU)                                           &                                            & 0.0994$^{+0.0030}_{-0.0032}$         \\ 
        $i$ (deg)                                          &                                            & 86.9$^{+1.9}_{-1.7}$                 \\ 
        $R_\mathrm{p}$ ($R_\mathrm{J}$)                  &                                              & 1.058$^{+0.068}_{-0.062}$\textasteriskcentered \\ 
        $M_\mathrm{p}$ ($M_\mathrm{J}$)                  &                                              & 1.49$\pm$0.13                        \\ 
        $\rho_\mathrm{p}$ (g$\cdot$cm$^{-3}$)              &                                            & 1.74$^{+0.19}_{-0.18}$               \\ 
        \bottomrule
        \bottomrule
        \multicolumn{3}{l}{
            \textsuperscript{\textasteriskcentered} See the end of Section~\ref{sec:results} for a revised value of the upper uncertainty.
        } \\
    \end{tabular}
    \label{tab:tic204_joint_params}
\end{table}

%% file: tables/tic394_rv_obs.tex
\begin{table}
    \centering
    \setlength{\tabcolsep}{10pt}
    \renewcommand{\arraystretch}{1.2}
    \caption{Radial Velocity Observations for TOI-4377}
    \begin{tabular}{llll}
        \toprule
        \toprule
        Time (BTJD) & RV (m$\cdot$s$^{-1}$) & $\sigma_\mathrm{RV}$ (m$\cdot$s$^{-1}$) & Instrument \\
        \midrule
        2319.72448 & 12562.68 & 37.77 & CORALIE \\ 
        2322.68777 & 12584.39 & 30.71 & CORALIE \\ 
        2327.64425 & 12621.87 & 54.49 & CORALIE \\ 
        2328.58199 & 12631.70 & 60.34 & CORALIE \\ 
        2329.72347 & 12434.57 & 38.66 & CORALIE \\ 
        2331.65101 & 12549.34 & 37.61 & CORALIE \\ 
        2333.65451 & 12474.46 & 29.82 & CORALIE \\ 
        2335.64513 & 12570.10 & 37.95 & CORALIE \\ 
        2343.52434 & 12460.12 & 42.84 & CORALIE \\ 
        2345.57824 & 12644.60 & 26.34 & CORALIE \\ 
        2346.66669 & 12546.54 & 42.58 & CORALIE \\ 
        2357.59247 & 12630.90 & 36.86 & CORALIE \\ 
        2358.55233 & 12645.13 & 42.20 & CORALIE \\ 
        2361.50828 & 12501.66 & 21.45 & CORALIE \\ 
        2365.56857 & 12492.95 & 18.46 & CORALIE \\ 
        2369.52124 & 12460.29 & 19.62 & CORALIE \\ 
        2371.53820 & 12667.72 & 24.29 & CORALIE \\ 
        2372.65545 & 12587.73 & 30.72 & CORALIE \\ 
        2375.51523 & 12660.14 & 26.66 & CORALIE \\ 
        2377.52719 & 12524.64 & 22.04 & CORALIE \\ 
        2349.59669 & 11163.26 & 42.04 & CHIRON \\ 
        2360.52207 & 10927.55 & 10.90 & CHIRON \\ 
        2364.51929 & 11008.63 & 14.42 & CHIRON \\ 
        2366.48316 & 11094.18 & 9.82 & CHIRON \\ 
        2367.50442 & 11130.68 & 10.20 & CHIRON \\ 
        2374.50064 & 11002.46 & 11.35 & CHIRON \\ 
        \bottomrule
    \end{tabular}
    \label{tab:tic394_rv_obs}
\end{table}

%% file: tables/tic204_rv_obs.tex
\begin{table}
    \centering
    \setlength{\tabcolsep}{10pt}
    \renewcommand{\arraystretch}{1.2}
    \caption{Radial Velocity Observations for TOI-4551}
    \begin{tabular}{llll}
        \toprule
        \toprule
        Time (BTJD) & RV (m$\cdot$s$^{-1}$) & $\sigma_\mathrm{RV}$ (m$\cdot$s$^{-1}$) & Instrument \\
        \midrule
        2344.58473 & 41184.15 & 8.32 & CHIRON \\ 
        2349.63372 & 41435.71 & 5.65 & CHIRON \\ 
        2356.51315 & 41282.27 & 5.77 & CHIRON \\ 
        2359.59827 & 41418.42 & 4.72 & CHIRON \\ 
        2378.57253 & 41406.44 & 5.92 & CHIRON \\ 
        2382.54754 & 41231.28 & 9.57 & CHIRON \\ 
        2383.46308 & 41170.69 & 7.85 & CHIRON \\ 
        2385.61740 & 41239.41 & 7.54 & CHIRON \\ 
        2387.61159 & 41345.90 & 8.80 & CHIRON \\ 
        2409.52933 & 41370.92 & 6.28 & CHIRON \\ 
        2416.50498 & 41298.20 & 7.17 & CHIRON \\ 
        2418.53099 & 41368.87 & 5.78 & CHIRON \\ 
        2419.50210 & 41382.79 & 6.16 & CHIRON \\ 
        2425.48791 & 41199.13 & 5.66 & CHIRON \\ 
        2430.49159 & 41364.45 & 5.37 & CHIRON \\ 
        2395.76900 & -102.03 & 1.50 & HIRES \\ 
        2399.76400 & 66.96 & 1.70 & HIRES \\ 
        2654.89700 & -25.20 & 2.10 & HIRES \\ 
        2681.82100 & -79.71 & 2.20 & HIRES \\ 
        2694.97700 & -25.30 & 2.20 & HIRES \\ 
        2724.81600 & -21.00 & 2.20 & HIRES \\ 
        2745.78800 & 55.40 & 2.10 & HIRES \\ 
        2748.77700 & 130.50 & 1.80 & HIRES \\ 
        \bottomrule
    \end{tabular}
    \label{tab:tic204_rv_obs}
\end{table}

%% file: tables/symbol_names.tex
\begin{table}
    \centering
    \setlength{\tabcolsep}{6pt}
    \renewcommand{\arraystretch}{1.1}
    \caption{Parameters shown in Tables~\ref{tab:tic394_joint_params} and \ref{tab:tic204_joint_params}}
    \begin{tabular}{ll}
        \toprule
        \toprule
        Symbol                                          & Definition    \\
        \midrule
        $P$ (days)                                      & Orbital period \\ 
        $t_0$ (BTJD)                                    & Time of inferior conjuction \\ 
        $b$                                             & Impact parameter \\ 
        $u_{1,2}$                                       & Quadratic limb-darkening coefficients \\  
        $\log\sigma_\mathrm{TESS}$ (ppm)                & Photometric shot noise (TESS dataset) \\
        $a_\mathrm{gran}$ (ppm)                         & Granulation amplitude parameter \\ 
        $b_\mathrm{gran}$ ($\mu$Hz)                     & Granulation timescale parameter \\ 
        $K$ (m$\cdot$s$^{-1}$)                          & Radial velocity semi-amplitude \\ 
        $\gamma_\mathrm{CHIRON}$ (m$\cdot$s$^{-1}$)     & Systemic velocity (CHIRON)  \\ 
        $\gamma_\mathrm{HIRES}$ (m$\cdot$s$^{-1}$)      & Systemic velocity (HIRES) \\ 
        $\gamma_\mathrm{CORALIE}$ (m$\cdot$s$^{-1}$)    & Systemic velocity (CORALIE) \\ 
        $\sigma_\mathrm{CHIRON}$ (m$\cdot$s$^{-1}$)     & Radial velocity jitter (CHIRON) \\ 
        $\sigma_\mathrm{HIRES}$ (m$\cdot$s$^{-1}$)      & Radial velocity jitter (HIRES) \\ 
        $\sigma_\mathrm{CORALIE}$ (m$\cdot$s$^{-1}$)    & Radial velocity jitter (CORALIE) \\ 
        $e$                                             & Orbital eccentricity \\ 
        $\omega$ (deg)                                  & Longitude of periastron \\ 
        $a$ (AU)                                        & Semi-major axis \\ 
        $i$ (deg)                                       & Orbital inclination \\ 
        $R_\mathrm{p}$ ($R_\mathrm{Jup}$)               & Planetary radius (in Jupiter radii) \\ 
        $R_\star$ ($\mathrm{R}_\odot$)                  & Stellar radius (in solar radii) \\ 
        $M_\mathrm{p}$ ($M_\mathrm{Jup}$)               & Planetary mass (in Jupiter masses) \\ 
        $M_\star$ ($\mathrm{M}_\odot$)                  & Stellar mass (in solar masses) \\ 
        $\rho_\mathrm{p}$ (g$\cdot$cm$^{-3}$)           & Planetary mean density \\
        \bottomrule
    \end{tabular}
    \label{tab:symbol_names}
\end{table}